\documentclass{article} 
\usepackage{aaai25}  
\usepackage{times}  
\usepackage{helvet}  
\usepackage{courier}  
\usepackage[hyphens]{url}  
\usepackage{graphicx} 
\usepackage{natbib}  
\usepackage{caption} 
\usepackage{booktabs}
\usepackage{multirow}
\usepackage{algorithm}
\usepackage{algorithmic}

\usepackage{newfloat}
\usepackage{listings}
\DeclareCaptionStyle{ruled}{labelfont=normalfont,labelsep=colon,strut=off} 
\floatstyle{ruled}
\newfloat{listing}{tb}{lst}{}
\floatname{listing}{Listing}
 \nocopyright 

\title{LLM-ABM for Transportation: Assessing the Potential of LLM Agents in System Analysis}
\author{
    Tianming Liu\textsuperscript{\rm 1},
    Jirong Yang\textsuperscript{\rm 2},
    Yafeng Yin\textsuperscript{\rm 1}
}
\affiliations{
    \textsuperscript{\rm 1}Department of Civil and Environmental Engineering, University of Michigan\\
    \textsuperscript{\rm 2}Department of Computer Science and Engineering, University of Michigan\\

}

\usepackage{bibentry}

\begin{document}

\maketitle

\begin{abstract}
Agent-based modeling approaches represent the state-of-art in modeling travel demand and transportation system dynamics and are valuable tools for transportation planning. However, established agent-based approaches in transportation rely on multi-hierarchical mathematical models to simulate travel behavior, which faces theoretical and practical limitations. The advent of large language models (LLM) provides a new opportunity to refine agent-based modeling in transportation. LLM agents, which have impressive reasoning and planning abilities, can serve as a proxy of human travelers and be integrated into the modeling framework. However, despite evidence of their behavioral soundness, no existing studies have assessed the impact and validity of LLM-agent-based simulations from a system perspective in transportation. This paper aims to address this issue by designing and integrating LLM agents with human-traveler-like characteristics into a simulation of a transportation system and assessing its performance based on existing benchmarks. Using the classical transportation setting of the morning commute, we find that not only do the agents exhibit fine behavioral soundness, but also produce system dynamics that align well with standard benchmarks. Our analysis first verifies the effectiveness and potential of LLM-agent-based modeling for transportation planning on the system level.
\end{abstract}

%

\section{Introduction}
In planning urban transportation infrastructure systems, modeling the travel demand of residents is a critical task in the planning process \cite{meyer2016transportation}. The travel demand model aims to capture the travel-related choices and decisions of travelers based on a series of social and economic factors including demographics, built environment, economy, and transportation infrastructure, and is used to predict the travel patterns in the transportation system and determine the interaction between travelers and transportation infrastructure in order to obtain system performance. Currently, the state-of-the-art travel demand models are the agent-based models and the corresponding microsimulation tools \cite{huang2022overview}. Compared to the traditional four-step model of travel demand \cite{mcnally2007four}, the agent-based models 
avoid trip-level aggregation biases by modeling travelers with representative agents who can interact with the environment and make informed, autonomous decisions \cite{bernhardt2007agent}. In transportation engineering practice, agent-based models have also begun to replace the classical four-step models in the planning process of some planning agencies \cite{wingo2023snapshot}.

However, the established agent-based transportation demand modeling methods still have some limitations that limit their behavioral realism and applicability in practice. First, the decision-making mechanisms within agent-based models generally rely on mathematical models that require \textit{a priori} behavioral assumptions, which may not fully capture the bounded rationality and nuanced decision-making processes of travelers \cite{ben1985discrete,van2022choice}. Second, these models often demand extensive, high-quality local data for calibration, particularly for multi-modal systems, which poses a significant barrier to widespread application in real-world scenarios \cite{manzo2014potentialities,kagho2020agent}. Additionally, evaluating alternative plans and scenarios, particularly the impact of complex, targeted policies or emerging technologies within the current framework, often demands extensive editing and coding \cite{bucci2018fhwa,wingo2023snapshot}. This places significant requirements on both the design of the modeling framework and the expertise of the personnel conducting the evaluations, which poses additional challenges for its usage in planning agencies.

In recent years, Large Language Models (LLMs) \cite{ouyang2022traininglanguagemodelsfollow} have exhibited exceptional reasoning and planning capabilities, closely aligning with human cognition by gathering and inferring information from natural language. Therefore, LLMs have been widely employed as agents in various frameworks to simulate human behavior within complex and interactive systems \cite{guo2024largelanguagemodelbased,Wang_2024,xi2023risepotentiallargelanguage}. In the field of transportation, extensive research has employed LLMs to generate travel data and draft travel plans \cite{chen2024travelagent, liu2024can,liu2024human, wang2024largelanguagemodelsurban} and showed LLMs have the potential to represent complex behavior of humans and has less calibrated data requirement. However, these established researches focus on behavior simulation at a single time point without considering traveler-infrastructure interaction, and LLMs still lack evaluation within the context of transportation systems dynamics and performance analysis, limiting their cogency in transportation planning applications. We believe it is necessary to systematically analyze LLMs within the context of transportation, as such analysis can elucidate the capability boundaries of LLMs and offer guidance for their broader application in social science research. However, applying LLMs in such real-world simulations inevitably makes it challenging to evaluate the validity of experimental results due to the open-ended nature of scenarios and outcomes \cite{10.1145/3571730,xu2023exploring}. Current methods mostly rely on historical data 
\cite{chopra2024limits}, human evaluations 
\cite{park2023generative}, and the self-consistency of LLMs \cite{nascimento2023self} to support experimental results. In this paper, we utilize a well-known and accepted case in transportation engineering: the morning commute, which has well-known benchmarks for system outcomes, to validate our LLM simulation results.


Importantly, in transportation systems, direct interactions between individuals are rare. Instead, the collective behavior of travelers influences the performance of transportation infrastructure, which in turn affects each individual's travel decisions. However, this differs from the classic imperfect information games in game theory. Although individuals are aware of these dynamics, they do not always engage in rational strategic decision-making but retain certain behavioral inertia, making decisions under bounded rationality \cite{di2016boundedly,ye2017rational,yu2020day}. This phenomenon is not unique to the field of transportation \cite{chen2012modeling,bischi2000global,samuelson1995bounded}. Through this paper, we hope to provide insights for researchers who use agents to simulate the real world in future studies.

\begin{figure}[t]
\centering
\includegraphics[width=0.9\columnwidth]{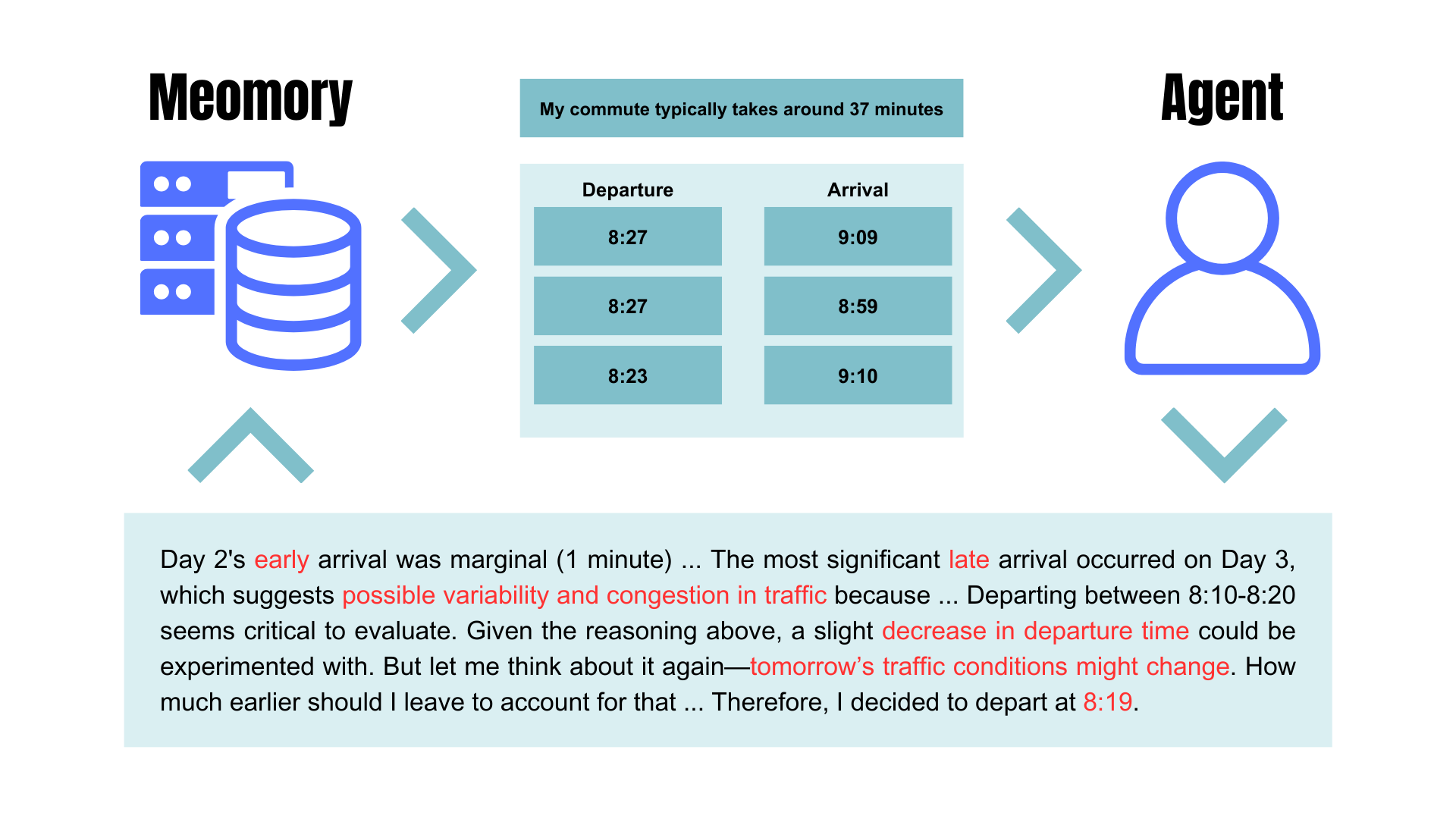} 
\caption{Illustration of the LLM agent perception and decision-making pipeline}
\label{fig:overall_map}
\end{figure}

In this paper, we integrate LLM agents into an agent-based modeling framework of transportation systems. Each agent serves as a proxy for human travelers and relies solely on its own historical data and travel experiences to make decisions, as shown in Figure \ref{fig:overall_map}. Furthermore, we also design the agents with a memory system,  reflection abilities, and bounded rationality to increase their similarity with human travelers. To model the transportation system dynamics, which is a multi-agent competition game without full information, we combine the LLM agents with a dynamic traffic assignment simulation of transportation infrastructure and conduct tests on the classical morning commute scenario. Our results not only demonstrate the behavioral soundness of LLM agents but also illustrate that the system outcome generated by the LLM-agent-based simulation also eventually converges closely with theoretical benchmarks in transportation engineering. Our findings shed light on the positive values of LLM-agent-based modeling in the evaluation and planning of transportation systems.

The contributions of this work can be summarized as follows:

\begin{itemize}
    \item We take the first step to integrate LLMs into the domain of agent-based modeling of transportation systems for transportation planning applications, advancing the fusion of LLMs and transportation research at a finer granularity.
    \item We introduce imperfect information games under zero-communication and bounded rationality of agents into LLM simulations for the first time, identifying scenarios that established works have not considered in real-world simulations.
    \item By combining theoretical derivations and micro-simulations, we validate our experimental results against the common benchmarks of system outcomes in transportation modeling and planning, offering valuable insights for future research in this area.
\end{itemize}
\section{Related work}
\subsection{Agent-based Models in Transportation}
In agent-based models of transportation systems, the agents represent individual human travelers in the transportation system. In established approaches, the travel behavior of human travelers is captured by a set of rules that govern the learning and decision-making process of the agents. By simulating agents’ detailed travel behaviors and incorporating these behaviors into a dynamic representation of both demand and supply in transportation systems, agent-based models provide detailed, granular insights that improve forecast accuracy. Currently, a diverse set of agent-based models such as TRANSSIMS \cite{smith1995transims}, MATSIM \cite{w2016multi}, SIMMOBILITY \cite{adnan2016simmobility} and POLARIS \cite{auld2016polaris} have been developed by transportation researchers for transportation system evaluation and planning.

\subsection{LLM Agents in Human-Behavior Simulations}
Recently, with advancements in thought mechanisms and improvements to the models, LLMs have demonstrated impressive reasoning and decision-making capabilities. Researchers leverage LLMs to simulate human behavior in specific settings by enabling communication both among multiple agents and between agents and their environment, which allows for a deeper understanding of phenomena or more reasonable forecasts. Therefore, LLM agents have been integrated into various frameworks of real-world simulation, such as social networks \cite{gao2023s, papachristou2024network}, small sandbox social systems \cite{park2023generative,wang2023humanoidagentsplatformsimulating}, economic systems \cite{han2023guinea,li2024econagent,weiss2024redesigning}, public health systems \cite{williams2023epidemic, chopra2024limits}, cooperative task solving \cite{zhang2023exploring,chen2023agentverse}, and travel plans drafting \cite{chen2024travelagent,tang2024synergizing}, etc. 

These studies have demonstrated the versatility and potential of LLM agents in simulations. However, most simulations are conducted in open-ended environments, with few studies validating the results of LLM agents' simulations in the context of their fields. Additionally, unlike previous applications, most scenarios in transportation systems, including the travel demand modeling explored in this paper, are imperfect-information, non-zero-sum games, where individuals operate under bounded rationality and have no communication with one another.
\section{Framework}
\subsection{Preliminaries}
To capture the human travel behavior with sufficient resolution for the analysis, the agent's decision-making process usually captures three main behaviors:
\begin{itemize}
    \item Activity schedule: the first decision captured by the agent framework is travelers' daily activity schedules. Agent $i$ will determine its activity schedule $a_i^t$ at day $t$, which includes its overall schedule of activities, the activity types, locations, and the travel departure time to go to these activities. 
    \item Mode choice: the second decision captured by the agents is the travel mode choice of the agents. For each segment of travel, agent $i$ will determine the travel mode $m_i^t$ (e.g. driving, biking, taking a bus) it would take.
    \item Route choice: the final decision captured by the agents is the travel route choice of the agents. For each segment of travel, agent $i$ will determine the route $r_i^t$ it should take in their travel.
\end{itemize}

It is worth noting that the activity scheduling decision has a continuous decision space while the latter two decisions are within a discrete space.

When making travel decisions, the agent will consider past travel experiences and their own social-demographic identity to make the decisions. In established models, the past travel experience is commonly represented by the experience $y_i^{t-1}$ in the previous day $t-1$. This information, combined with the agent's social-demographic characteristics and living environment $s_i$, jointly determines the agent's behavior through a comprehensive mathematical model $f$:
\begin{equation}
    a_i^t,m_i^t,r_i^t=f(y_i^{t-1},s_i)
    \label{eq:agent_decision}
\end{equation}

The mathematical model $f$ applied in current agent-based frameworks These models all require \textit{a priori} behavioral assumptions and can be roughly categorized into two categories: econometric-based and rule-based computational process models. Econometric models, such as the discrete choice model in Simmobility \cite{adnan2016simmobility}, rely on discrete choice theory to capture travelers' activity patterns assuming that travelers are utility maximizers. However, these assumptions overlook human's bounded rationality, limiting the models' ability to fully capture human travel behavior. Rule-based models, like the classification tree in TRANSSIMS \cite{smith1995transims} and the candidate travel plan selection method in MATSim \cite{w2016multi}, use predefined rules to simulate activities and destinations. While less computationally demanding, these models struggle with the accuracy and generalization of their heuristics and face challenges in fully representing travelers' behavior as the rules are fixed in different situations. Furthermore, existing frameworks also require extensive data to be properly calibrated. This barrier significantly affects the accessibility of advanced agent-based models to transportation planners and practitioners. Per a report in 2015 \citep{systematics2015status}, only 16\% of the metropolitan planning agencies in the United States have plans to move from conventional four-step models to more advanced models in the foreseeable future.

After the agents have made their travel decisions, the  agents' realized travel results $y_i^t$ are determined by a simulator $g$:
\begin{equation}
    y_i^t=g(a_i^t,m_i^t,r_i^t,a_{-i}^t,m_{-i}^t,r_{-i}^t)
\end{equation}
In practice, this simulator is often a comprehensive dynamic traffic network simulator \cite{barcelo2005dynamic,zhou2014dtalite} that simulates the traffic flows on the transportation network and the corresponding links.

In this paper, we focus our analysis on the morning commute situation, which is a classic case for model development and testing in transportation engineering. The morning commute case contains varied behavioral tasks for the agents, including deciding their travel schedule and route choice, as well as  The morning commute cases not only offer both continuous and discrete decision problems but also have widely accepted benchmarks traffic patterns for evaluation of our simulation results \cite{arnott1990economics,smith1984stability}. Thus, we use the corresponding cases to design our simulation experiments.

\subsection{Agent behavior}
\begin{figure}[h]
\centering
\includegraphics[width=0.9\columnwidth]{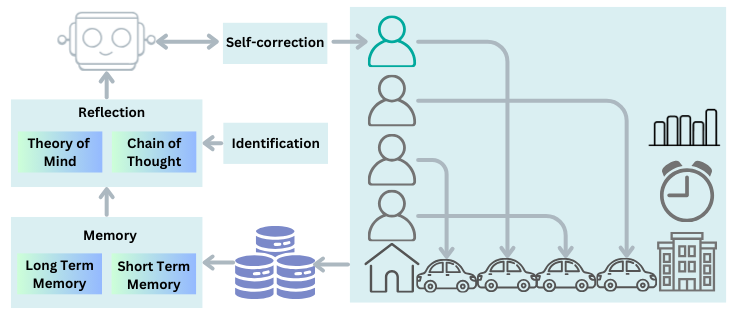} 
\caption{Behavioral pipeline of LLM traveler agents}
\label{fig:agent_design}
\end{figure}

Our overall design of agent behavior is illustrated in Figure \ref{fig:agent_design}.

Each agent determines its travel decision based on its memory system and planning mechanisms, as outlined below:
\subsubsection{Memory}
\begin{itemize}
   \item Long-Term Memory: For each OD (origin-destination) pair, the agent maintains a history of past travel experiences. Alongside this data, the agent retains a fuzzy impression of typical commute times under normal conditions, providing a general reference for departure planning.
   \item Short-Term Memory: The agent extracts recent travel data from the memory pool, typically covering the past K days, as its short-term memory. Given the day-to-day variations in traffic conditions, short-term memory focuses more on capturing recent anomalies and trends compared to the broader perspective provided by long-term memory.
\end{itemize}
\subsubsection{Reflection}
To facilitate more sophisticated analyses of memory and departure planning, as well as more realistic simulation of the behavior of travelers, the following mechanisms are integrated into the prompt:
\begin{itemize}
\item Chain of Thought (CoT)\cite{wei2023chainofthoughtpromptingelicitsreasoning}:
This mechanism facilitates step-by-step reasoning, helping the agent analyze key insights from short-term memory, such as the average travel time and its variability over the past few days, as well as patterns of early or late arrivals.
\item Theory of Mind (ToM)\cite{Strachan2024TestingTO}:
Under the theory of mind, agents are aware that the system involves multiple participants but do not engage in detailed simulations of other agents’ behavior, as would be done in imperfect-information zero-sum games. Instead, this mechanism enables agents to recognize that while they adjust their departure times, other agents are doing the same, leading to shifts in peak periods. By accounting for this, agents can avoid falling into a vicious cycle of “depart earlier—encounter congestion—depart even earlier—repeat.”
\item Bounded Rationality:
In modern psychology and behavioral economics theory, humans exhibit strongly bounded rationalities in their decision-making \cite{kahneman2003maps}. Humans may exhibit biases, loss aversion, risk aversion, and behavioral inertia (unwillingness to change from the status quo) in their decision-making process that prevents them from making the most rational decision. To allow for more realistic simulation, we also incorporate bounded rationality into agents using prompt engineering.
\item Self-Correction\cite{kamoi2024llmsactuallycorrectmistakes}:
After making an initial decision, the agent simulates the potential outcomes of this decision for the day. By reflecting on these outcomes, the agent can refine its decision-making process and improve future planning strategies.
\end{itemize}
\subsection{Traffic Flow Simulation}
Once all agents decide on their travel choice for the day, traffic flow simulation is carried out using the Dynamic Traffic Assignment (DTA) method \cite{zhou2014dtalite}. The simulation proceeds as follows:
\begin{itemize}
\item Input Data:
The departure times and OD pairs of all agents, along with road network parameters, are fed into the DTA model. 
\item Simulation Process:
The DTA algorithm dynamically assigns traffic flows to each road segment based on traffic demand and road network capacity. This process simulates vehicle movements across the network, generating road-level traffic conditions such as speed, flow, and congestion, while updating travel times for each route.
\item Output Data:
At the end of the simulation, the DTA algorithm computes the actual arrival times for each agent. This data is then recorded to update the agents’ short-term memory. 

Through DTA, we can quantify how individual behaviors influence collective traffic flow and further explore the feedback effects of collective behavior on individual behavior.
\end{itemize}
\section{Experiments}
\subsection{Continuous choice case: commute departure time choice}
\subsubsection{Setup}
The first case in our analysis is the bottleneck morning commute problem, in which agents would choose the desired departure time to go to work when facing congestion in their commute. All agents have a corridor of one single route connecting a uniform origin and destination, and there is a bottleneck that has a limited capacity. The agents may form a point queue and face congestion at the bottleneck location. All agents have a work start time of 9:00 am and they seek to balance the disutility of early arrival, in-route travel time, and late arrival. The ratio of the marginal disutility of early arrival time, in-route travel time, and late arrival time is 1:3:10.

In our case study, there are 40 agents commuting using the corridor. The corridor has a free-flow (base) travel time of 30 minutes, and the capacity of the bottleneck is 60 vehicles per hour. In transportation literature, a common benchmark of the arrival time pattern is a uniform distribution between 8:24 am and 9:04 am.

\textbf{Agents behavior setting}
Each agent aims to minimize their marginal disutility by balancing early arrival time, late arrival time, and travel time. Two main features influence their decisions: they draw on memories of past departure and arrival times to infer today’s departure time, and they exhibit bounded rationality with behavioral inertia when deciding whether to adjust their departure time. In our simulation, 40 agents repeatedly make departure time choices for 40 days in this case.

\subsubsection{Result and analysis}
\begin{table}[h!]
\centering
\renewcommand{\arraystretch}{1.2} 
\begin{tabular}{@{}lcccc@{}}
\toprule
\multirow{2}{*}{Time Interval} & \multicolumn{4}{c}{Days in Simulation} \\ 
\cmidrule(lr){2-5}
                               & 1-10 & 11-20 & 21-30 & 31-40 \\ 
\midrule
\textit{Departure time } & & & &\\
\textit{distribution}& & & &\\
7:45-8:00                     & 0.25\%       &  0.50\%       & 1.25\%      & 1.01\%       \\
8:00-8:15                     & 18.32\%   & 80.00\%   & 96.75\%     & 97.49\%   \\
8:15-8:30                     & 74.81\%   & 19.50\%    & 2.00\%    &  1.51\%   \\
8:30-8:45                     & 6.62\%    & 0.00\%     & 0.00\%    & 0.00\%    \\
\textit{Arrival time } & & & &\\
\textit{distribution}& & & &\\
8:30-8:45*                    & 11.53\%    & 23.00\%    & 36.00\%    & 39.50\%    \\
8:45-9:00*                    & 45.61\%  & 41.50\%    & 43.75\%   & 43.75\%   \\
9:00-9:15*                    & 34.84\%   & 30.25\%   & 18.25\%   & 15.50\%    \\
9:15-9:30*                    & 7.77\%   & 4.75\%    & 0.75\%    & 0.50\%     \\
\bottomrule
\end{tabular}
\caption{Distribution of agent departure and arrival times.}
\label{tab:travel_distribution}
\end{table}
\begin{figure}[t]
\centering
\includegraphics[width=0.49\columnwidth]{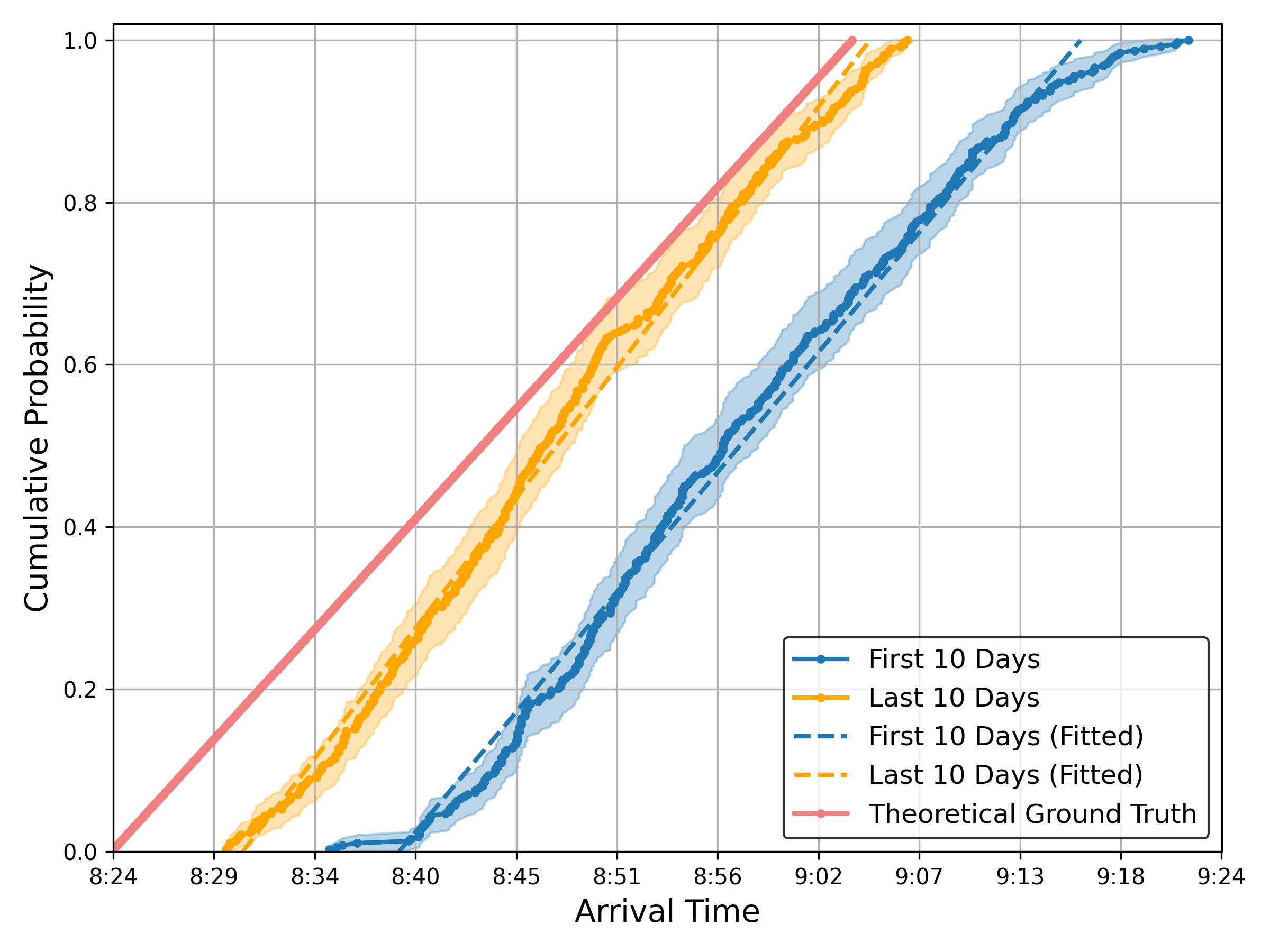} 
\includegraphics[width=0.49\columnwidth]{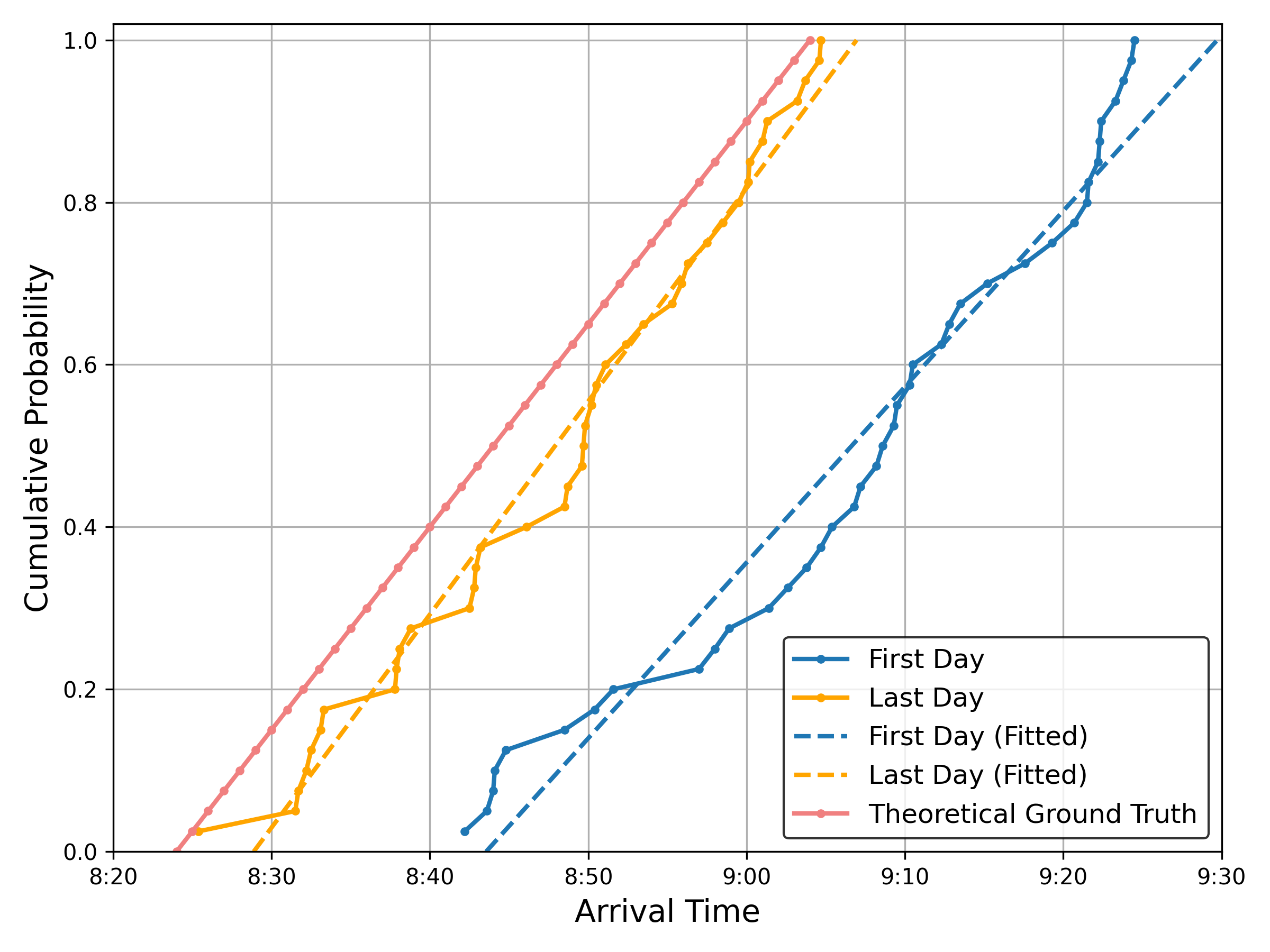} 
\caption{Distribution of Arrival Times: First Day(Group) vs. Last Day(Group)}
\end{figure}
\begin{figure}[t]
\centering
\includegraphics[width=0.9\columnwidth]{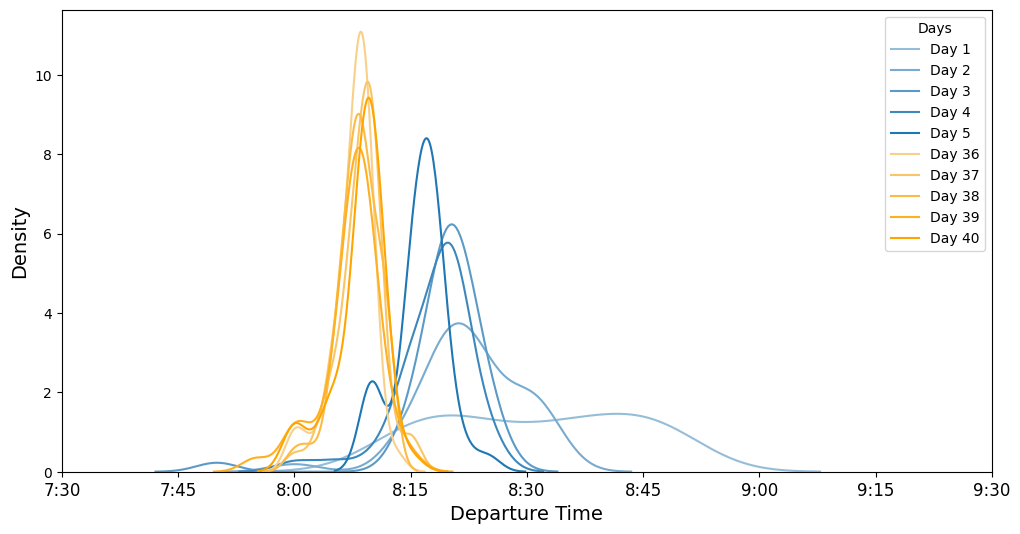} 
\caption{Distribution of Departure Times: First 5 Days vs. Last 5 Days}
\label{fig1}
\end{figure}
The experimental results are presented in Table 1. To highlight trends over a consistent period and reduce the impact of minor daily fluctuations, the data is grouped into 10-day intervals for analysis. As the simulation progresses, the departure times of multiple agents become increasingly concentrated, with 97.49\% falling within the 8:00–8:15 range. This shift in departure times also leads to earlier arrivals, reducing the tardiness rate from over 40\% to 16\%. As illustrated in Figure 1, departure times transition from an initially disordered distribution in the early days to a final state of convergence, resembling a normal distribution. Meanwhile, arrival times are approximately uniformly distributed between 8:30 am and 9:07 am. Our simulation's arrival time distribution is very close to the benchmark uniform distribution between 8:24 am and 9:04 am.

To further analyze the ability of LLMs to mimic human decision-making, we investigated their behavioral inertia. In the persona module, we limited their capacity to adjust departure times based on human characteristics, ensuring they do not make significant changes unless necessary. While this condition could be defined more specifically—such as specifying the severity of traffic congestion experienced or the length of tolerated lateness—a vague constraint was sufficient for the LLM to comprehend. Conversely, in the reasoning component, we prompted that if it encounters traffic congestion and wishes to adjust its time, minor adjustments would not alleviate the situation, so larger changes are required. Although these two prompts are opposite, they reflect the real-life contradictions humans face in such scenarios. In our experiments, we observed that when the LLM became aware of traffic congestion, it sometimes failed to adjust its timing successfully. This also helps agents with identical settings make different choices when facing the same situation.

Additionally, we conducted a brief ablation study. Without the theory of mind module guiding agents to consider changes in peak times, they tend to fall into a cycle of departing earlier, encountering congestion, departing even earlier, and repeating the process. This results in departure times being concentrated in relatively early periods and longer average travel times. Without the behavioral inertia module, agents become overly sensitive, making large adjustments and leaving excessive time buffers in response to congestion or delays. This leads to slower convergence or, in some cases, failure to converge altogether.
\subsection{Discrete choice case: commute route choice}
\subsubsection{Setup}
The second case in our analysis is the commuting route choice problem, in which the agents would choose one route to take from a finite set of available routes to commute to work. Without the loss of generality, we set up the transportation network so that all agents may choose one of two available routes in the network. Each route may have congestion that is dependent on the number of people $f_1$ and $f_2$ traveling on that route. The relationship between the route flows $f_1$, $f_2$ and the route travel times $t_1$, $t_2$ are as follows:
\begin{equation}
    t_1=20+3\times f_1
\end{equation}
\begin{equation}
    t_2=40+f_2
\end{equation}

Each agent aims to minimize their route travel time by choosing one route every day. Two main features will impact their decisions: they will use their memories of past route travel times to infer the travel time of the routes today, and they will also exhibit bounded rationality in behavior inertia when considering whether to change their route choice. Overall, in our simulation, 40 agents engage in repeated route choice for 20 days in this aforementioned network.

\subsubsection{Result and analysis}
\begin{figure}[t]
\centering
\includegraphics[width=0.9\columnwidth]{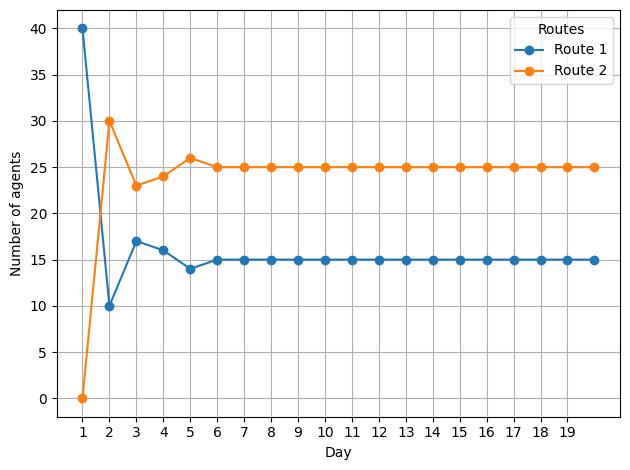} 
\caption{Progression of number of agents on each route}
\label{fig: dtd_route_flow}
\end{figure}

The overall progression of the number of agents choosing routes 1 and 2 during the 20-day period is shown in Figure \ref{fig: dtd_route_flow}.

Figure \ref{fig: dtd_route_flow} clearly demonstrates that the number of agents on each route quickly converges and stabilizes in our 20-day simulation. On day 1, without any prior information, all agents choose route 1 which has the smaller no-congestion travel time. After that, agents begin to explore other options, and some shift to route 2 while others stick to route 1 due to their behavioral inertia. After 6 days of adjustment, the number of agents on each route is $f_1=15$ and $f_2=25$, resulting in both routes having the same travel time $t_1=t_2=65$. With both route's travel times being equal, agents realize that switching routes would only increase the target route's congestion and thus result in a worse travel time for them. Thus, agents stay in their current route choice and the route flows stabilize over time.

Not only does the simulated route traffic flow converge, it also converges to the uniform standard of wardrop equilibrium that is widely applied in assessing transportation systems. This shows that the LLM-agent simulation can generate 

\section{Conclusion}
To our knowledge, this study represents the first research to both integrate LLM agents into the agent-based simulation of transportation systems and conduct evaluations of the LLM-agent-based approach on the system level. By leveraging the reasoning and planning mechanisms, we utilize LLMs to generate departure time from historical data, effectively simulating changes in travel demand resulting from agents' adjustment from past experiences and capturing the interaction between travelers and transportation infrastructure. Using the classical morning commute case study in transportation engineering, the validity of our agent and system design was assessed through comparisons of simulation results and established transportation theoretical benchmarks, demonstrating that LLM-based agents can be effectively applied in agent-based modeling for transportation systems. We believe this preliminary experiment and validation provide insight into the broader application of LLMs in transportation planning, especially in evaluating the performance of transportation plans.

However, as an early attempt to analyze LLMs in the context of transportation system modeling, our research has certain limitations. For instance, our ablation studies are not comprehensive. Additionally, we focused primarily on common cases, leaving more complex agents (e.g., different personas) and environments (e.g., multimodal transportation systems, network structures) unexplored.

In the future, we plan to employ LLM agents in more complex experimental settings and real-world scenarios to simulate travel demand. Our objective is to use LLM-based agents to better emulate human behavior, providing insights into how future transportation policies can be designed to better accommodate user needs.


\end{document}